\documentclass[10pt]{article}

\usepackage{times}
\usepackage{graphicx}
\usepackage[superscript,biblabel]{cite}
\usepackage{placeins}
\usepackage{scicite}

\usepackage{lineno}

\title{A Silent Build-up in Seismic Energy Precedes Slow Slip Failure in the Cascadia Subduction Zone} 

\author
{Claudia Hulbert$^{1,2}$, Bertrand Rouet-Leduc$^{2}$, Paul A. Johnson$^{2}$
\\
\normalsize{$^{1}$ Laboratoire de G\'eologie, D\'epartement de G\'eosciences, \'Ecole Normale Sup\'erieure, PSL Research University}, \\ 
\normalsize{CNRS UMR 8538, Paris, France}\\
\normalsize{$^{2}$Los Alamos National Laboratory, Geophysics Group, Los Alamos, New Mexico, USA}\\
\normalsize{$^\ast$C. Hulbert (email: chulbert@lanl.gov), B. Rouet-Leduc (bertrandrl@lanl.gov), and P. Johnson (paj@lanl.gov)}
}

\date{}

\begin{document} 


\baselineskip24pt


\maketitle


\begin{abstract}
We report on slow earthquakes in Northern Cascadia, and show that continuous seismic energy in the subduction zone follows specific patterns leading to failure. We rely on machine learning models to map characteristic energy signals from low-amplitude seismic waves to the timing of slow slip events.

We find that patterns in seismic energy follow the 14-month slow slip cycle. Our results point towards a recurrent build-up in seismic energy as the fault approaches failure. This behaviour shares a striking resemblance with our previous observations from slow slips in the laboratory. 
\end{abstract}

\pagebreak

Since their discovery in Japan at the turn of the millennium \cite{Obara2002,Obara2016}, slow earthquakes and associated tectonic tremor and low frequency earthquakes (LFEs) have been identified in most subduction zones as well as other tectonic environments \cite{Veedu2016, Obara2016,Peng2010,BerozaIde,Gomberg2010,Rubinstein2010,Rousseteaav3274}. Slow slip is transitional between the fast rupture of regular earthquakes and stable sliding along a fault interface \cite{Obara2002}, and occurs deep in faults, down-dip from the nucleation zone of damaging earthquakes at the transition from brittle to ductile zones \cite{scholz_2019}. In subduction tectonics, slow slip and tremor are thought to take place where temperatures drive dehydration of subducting material that increase pore pressures, inhibiting brittle failure \cite{Gao2017}. Slow earthquakes are characterized by short-term slip with durations of days or long-term slip with durations of months or years \cite{Obara2016}. The slowly slipping region is thought to denote the transition from unstable (seismogenic) to stable (creep) friction and therefore may define the depth limit of megathrust rupture \cite{Gao2017}. How the slow slip may couple to the locked region is an open and fundamental question \cite{Obara2016,scholz_2019,Frank2016}. Slow slip has been shown to occur before some large seismogenic ruptures in some instances, for example before the M9.0 Tohoku earthquake \cite{ITO201314} and the M8.2 Iquique earthquake \cite{Ruiz1165}, and has been observed preceding large earthquakes in other tectonic environments including some transform faults \cite{kuna2019,Rousseteaav3274}. These observations suggest that coupling between slow slip and large earthquakes occur in some instances but has not been systematically observed, perhaps due to limitations of geodetic measurements.

In laboratory studies of slow slip \cite{Hulbert2018} generated by a bi-axial shear device \cite{Marone1998,Kaproth2013,Scuderi2016}, we showed that the energy of continuous seismic waves follows characteristic patterns throughout the slip cycle, allowing us to estimate key properties of the laboratory fault (instantaneous fault displacement, friction, gouge layer thickness). In a first effort to generalize these results to Earth, the analysis of slow slip in Cascadia \cite{Rouet2018_b} revealed that statistical characteristics of continuous seismic signals encode the fault displacement rate, as measured by GPS. These characteristics are related to signal energy and resemble those identified in the laboratory. 

In the laboratory, continuous energy patterns also encode information regarding the future behavior of the fault (time remaining before the beginning and the end of the next slow slip event, duration and `magnitude' of the next slow slip event). Here, we seek to determine whether these same signatures can be used to infer the future behavior of the slowly slipping zone in Cascadia beneath Vancouver Island. 

\section*{Seismic data analysis and slow slip failure times in Cascadia}

\begin{figure}[ht!]
\begin{center}
\includegraphics[width=16cm,trim= 0 0 0 0]{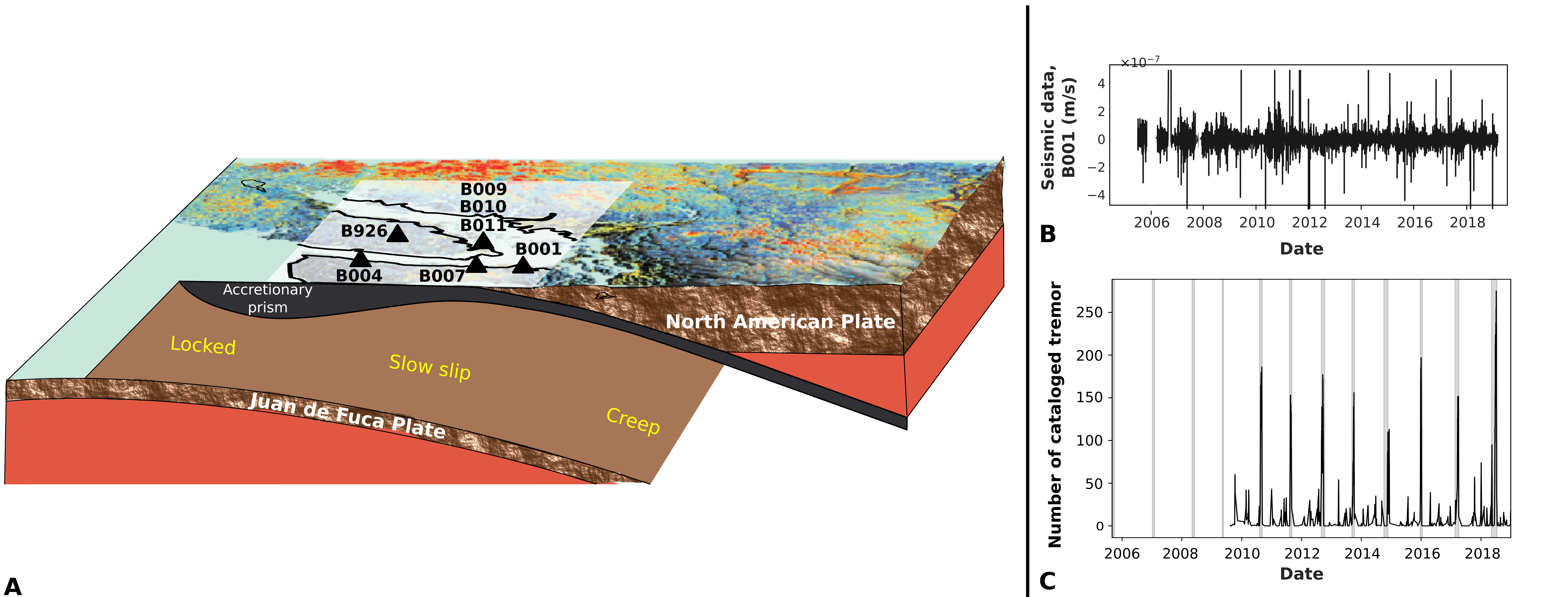}
\caption{\textbf{(A) Map of the area analyzed in Cascadia and sketch of the subduction zone.} The seismic stations are represented by black triangles.The shaded region represents the area of interest, the southern Victoria Island. Our goal is to rely exclusively on continuous seismic waves \textbf{(B)} to estimate the time remaining before the next slow slip event.  \textbf{(C)} Slow slip timing from PNSN Tremor Logs (gray shaded areas), and smoothed catalogued tremor rates from the PNSN catalog for comparison.} 
\label{fig:fig1}
\end{center}
\end{figure}

We analyze slow slip beneath Vancouver Island, Canada, where the Juan the Fuca oceanic plate subducts northwestly beneath the North American plate (Figure \ref{fig:fig1}). The Cascadia subduction zone exhibits slip events that are long-term, on the order of weeks \cite{Rogers2003,Gomberg2010,WechCreagerMelbourne2009,WechBartlow,Hawthorn2016,WechBartlow,Kao2005}. Large slow slip events occur quasi-periodically (approximately every 14 months), manifest by the North American plate lurching southwesterly over the Juan de Fuca plate. Smaller slow slip events occurring between these large periodic events have been identified recently, pointing towards a large variability in the size and timing of slow earthquakes in the area \cite{Frank2016}. We focus on Cascadia in scaling from the laboratory because of the recurrent slow slip events observed, especially in the vicinity of Vancouver Island \cite{Rogers2003}, and because there have been continuous seismic recordings for more than a decade. Supervised machine learning (ML) used in the work described below, requires robust training and testing sets with many slip events. The long history of slow slip observed in this region makes it an ideal case to determine if there is information carried in the seismic signal characteristic of upcoming failure. 

\FloatBarrier

ML analysis of seismic data is an expanding field, with many studies over the recent years focusing in particular on event detection \cite{Perol2018}, phase identification \cite{Ross2018}, phase association \cite{mcbrearty2019,Ross2019}, or patterns in seismicity \cite{Holtzman2018}. Here, by relying on a supervised ML approach, we assess whether continuous seismic waves carry the signature of how close the system is to failure. We use the PNSN Tremor Logs as a proxy for slow slip timing. These timings are estimated from bursts of tremor, and are represented by vertical gray bars in Figure \ref{fig:fig1} (C); tremor rates from the PNSN catalog \cite{WechBartlow} are also plotted for comparison. Note that because the tremor logs take into account the whole Cascadia region and are not geographically limited to Vancouver Island, slip timings may not match perfectly actual timings within the area of interest. We could use peaks in tremor rates within the area instead to identify failure times; the analysis below remains robust if we do so (see Supplementary). However, because the catalog only starts in 2009, this leads us to discard about a third of the data. We can also use GPS data to try to identify slow slip timing (see Supplementary).     

We rely on seven seismic stations from the Plate Boundary Observatory \cite{PBO}, denoted by black triangles in Figure \ref{fig:fig1} (A). We find that borehole stations are much more robust than surface ones regarding contamination by seasonal signals. A continuous, clipped and de-sampled seismic waveform for one of the stations is shown in Figure \ref{fig:fig1} (B). Our goal is to assess whether a given time window of this continuous seismic data can be used to estimate the time remaining until the next slow slip event. 

We use the methodology initially developed in laboratory shear experiments and modified in previous work \cite{Rouet2018_b} for application to Earth data. We first process the seismic data by correcting for the instrument gain, centering and de-trending the data, for each day during the period analyzed (2005-2018). The daily data intervals are band-passed between 8 and 13 Hz (within bands of 1Hz) and clipped, to limit the contribution of microseismic noise and earthquakes respectively, and to focus on low-amplitude signals. 
Once the data are pre-processed, for each day we compute a number of statistical features linked to signal energy (see Supplementary Information for details). These daily features are then averaged within a time window. Anomalous datapoints are detected within each window and removed before averaging using Isolation Forests \cite{Pedregosa2011}. The results shown below use a time window of 3 months, but our methodology is robust to changes in the window size (see Supplementary). Each window is indexed by its latest day; successive time windows are offset by one day. The averaged features over these time windows are used as input to the machine learning algorithm. In the following, `seismic features' will refer to those features averaged over a time window.

As in any supervised machine learning problem, our analysis is divided into a training and a testing portion. In the training phase, the algorithm takes as input the seismic features calculated from the first 50\% of the seismic data (training set), and attempts to find the best model that maps these features to the time remaining before the next slow slip event (label or target). The problem is posed in a regression setting. We select the model's hyper-parameters through Bayesian optimization by 5-fold cross-validation. We use Pearson's correlation coefficient (CC) as the evaluation metric.

Once a model is finalized, it is evaluated on data the model has never seen--the features from the remaining 50\% of the seismic data (testing set). It is important to note that in the testing phase, the model only has access to the seismic features calculated from the continuous seismic data, and has no information related to slow slip timing (the label). In the testing phase, the label is used exclusively to measure the quality of the model's estimates, \textit{i.e.} how close these estimates are compared to the true label values obtained from PNSN Tremor Logs. 

\section*{Estimating slip failure times from continuous seismic waves}

We rely on gradient boosted trees, algorithms that are transparent in their analysis in contrast to many other methods. These algorithms can be probed to identify which features are important in the model predictions, and why. Identifying the important statistical features allows us to compare with laboratory experiments, and gain insight into the underlying physics.

\begin{figure}[ht!]
\begin{center}
\includegraphics[width=16cm,trim= 0 0 0 0]{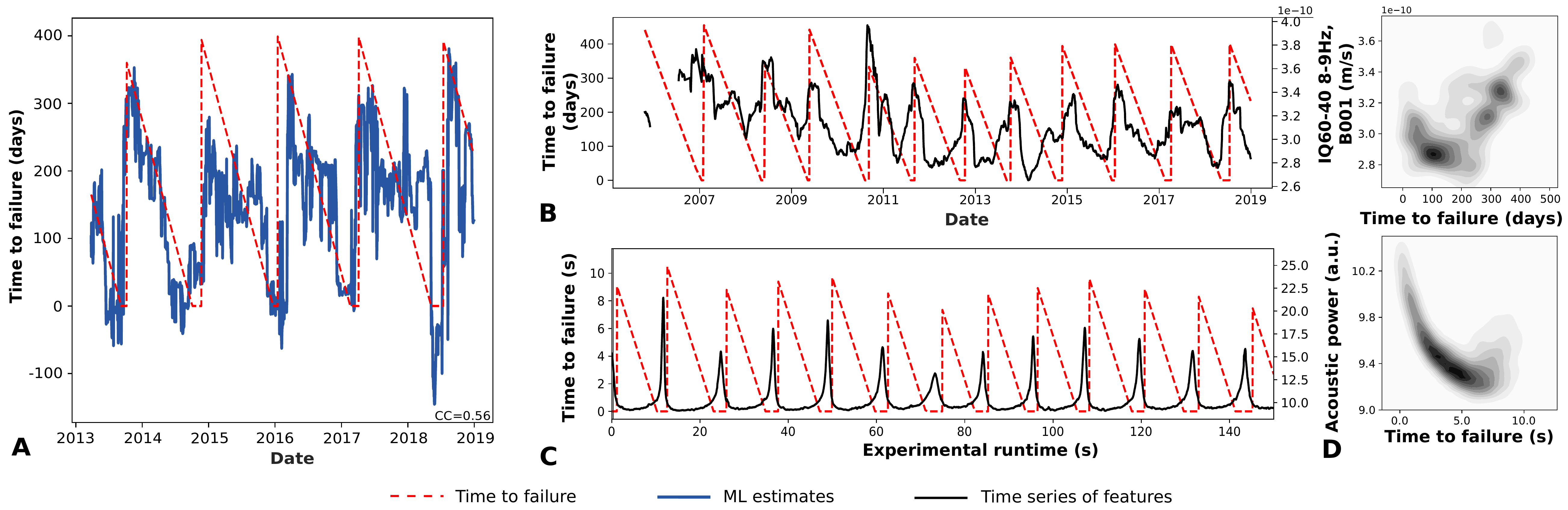}
\caption{\textbf{Slip timing estimations, patterns identified by our model, and comparison to shear experiments in the laboratory.} \textbf{(A)} ML estimates in testing of failure times (in blue), and time to failure (in red) from the PNSN tremor logs. \textbf{(B)} The most important feature identified by our model plotted against time, for the best stations (B001), for time window intervals of 3 months (black curve) [The right-hand vertical axes show the interquantile range (IQ 60-40) for the frequency band 8-9 Hz, and its evolution over a time window]. These energy-based features show clear patterns with respect to the time remaining before the next slow slip event (left axis).   \textbf{(C)} shows the best statistical feature found in laboratory slow slip experiments for comparison (acoustic power, right hand vertical axis). The best features in the laboratory (\textbf{C}) and Cascadia (\textbf{B}) are related to the energy of the seismic waves, and appear to follow very similar patterns, with a progressive increase in amplitude that peaks near the end of each slow slip event. \textbf{(D)} Distribution of these features with respect to time to failure, in between slip events - energy increases when failure approaches. Top: Cascadia slow earthquakes. Bottom: laboratory slow slips.
}
\label{fig:fig2}
\end{center}
\end{figure}

Estimations of the time remaining before the next slow slip event on the testing set are shown in Figure \ref{fig:fig2} (A). This plot shows the ML slip timing estimations (in blue) from station B001 (3-month windows) and the time remaining before the next slow slip event (ground truth, dashed red line). This ground truth can be understood as a countdown to the next slip, and is equal to zero whenever the PNSN reported an ongoing slow slip. The same method can also be applied successfully when using peaks in tremor rates or GPS to identify our proxies for failure times (see Supplementary).

When close to failure, the machine learning model is able to estimate the timing of most slow slip events - with the exception of the 2018 slow slip, which is not estimated before it begins. Each point on the blue prediction curve in Figure \ref{fig:fig2} (A) is derived from a single time window of seismic data. Thus, the results demonstrate that at any time during the slow slip cycle, a snapshot of continuous seismic waves is imprinted with fundamental information regarding the time remaining before the next slow slip event. In particular, the seismic data long before failure and close to failure appears very different to the trained model: although estimations far from failure are noisier, the model easily distinguishes between these two extreme cases in all the examples in our testing set. 

Our results suggest that the system follows specific patterns leading to failure. Note that because the events are most often separated by 13 or 14 months, the model cannot rely on seasonal signals to make its estimations. To prove this is the case, we show that a seasonal sinusoidal cannot lead to good estimations of failure times (see Supplementary). Therefore, the seismic data analyzed contains identifiable precursory patterns that match slow slip timing and are independent from the season. 

\FloatBarrier

\section*{Patterns in seismic energy follow the slow slip cycle, with a silent build-up preceding failure}

Because we choose to rely on transparent machine learning algorithms, we can identify the most important features used by our model to make its estimations of failure times, and therefore make comparison with laboratory experiments. We find that the best features identified in Cascadia follow almost identical patterns as those identified in the laboratory.

The most important seismic features identified by our model for the best station are plotted in Figure \ref{fig:fig2} (B). The evolution of these features over time follows clear 14-month patterns, and is related to the timing of the slow slip cycles - which explains why our model is able to make its estimations of slow slip timing. 

We know from recent work that similar continuous seismic features can be used to accurately estimate the displacement rate of the fault in both the laboratory \cite{rouet2017fault,Hulbert2018} and in Cascadia \cite{Rouet2018_b}. In particular, these statistical characteristics of the seismic data can be used to differentiate between the loading and slipping phases of the cycle. The estimations of failure times in Figure \ref{fig:fig2} (A) show that, just as in the laboratory, continuous seismic signals are also encoded with information regarding the future state of the fault. This suggests that some of the frictional physics scales from the simple laboratory system to subduction in Cascadia, which argues in favor of self-similarity of slow slip rupture and nucleation. 

In the laboratory, cycles in seismic energy seem to be associated to displacement on the fault. Therefore our results suggest that slow slips often begin with a silent build-up in energy possibly associated to an acceleration on the fault, that can be small enough to not be captured in the cataloged tremor - tremor at this early stage may be too weak to be picked up systematically. We find that a model trained only on cataloged tremor is not able to estimate slip timings (see Supplementary). In contrast, the peak energy measured during slow slips are likely driven by identified tremor events.

In the case of laboratory slow slip, the most important feature by far for forecasting failure time is the seismic power\cite{Hulbert2018}, shown in Figure \ref{fig:fig2} (C). In the case of the Cascadia subduction zone we rely on inter-quantile ranges, closely related to the root of the seismic power but with outlier values removed (these outliers are more likely to correspond to anthropogenic noise and/or local and distant earthquakes).

 Figure \ref{fig:fig2} (B), top, shows that seismic energy follows similar patterns in both the laboratory and Cascadia: i) a progressive increase, with peaks in energy reached toward the end of the slow slip (as the slipping phase of the cycle emits larger seismic amplitudes than the loading phase, due to episodes of tremor); and ii) an often abrupt decrease within each slip cycle, towards the end of an event. Cycles in Cascadia are clearly apparent, but much noisier. This is likely due to background noise and to the fact that many fault patches located over a broad spatial range on the subducting interface may be contributing to the slip. Nonetheless, the subduction zone at large scale follows similar patterns in energy; the machine learning algorithm identified these patterns, enabling timing estimations. 

In the laboratory, these signals carry fundamental information regarding the frictional state of the system. The fact that a similar behavior can be observed in Cascadia, with energy patterns matching the 14-month slow slip cycle, may also provide indirect information regarding the evolution of friction at much larger scale. The origin of the seismic signal is presumed to be due to fault gouge grain-to-grain and grain-to-block interactions. Effort is ongoing to explain its origin \cite{Gao2019,Ren2019}. In the Cascadia subduction zone, we posit that part of this energy is emitted from a large number of asperities located on the fault interface. 
To refine the analysis and focus exclusively on tremor, we are training models to recognize tremor and estimate cycles in tremor energy \cite{tremorness}, which may ultimately help to improve the slip timings estimations.

\section*{Conclusion}

We find that seismic energy in the Cascadia subduction zone follows specific patterns throughout the slow slip cycle, with energy starting to build up steadily before failure. This initial increase in energy is not captured in the cataloged tremor - suggesting that slow slip events often begin with a silent phase. Slow slip has been observed prior to large earthquakes in some instances, but not systematically. If indeed slow earthquakes have identifiable precursory signals, estimating their timing may ultimately help improve hazard assessment for destructive earthquakes. \\

\section*{Acknowledgements}
We thank C. Marone, C. Bolton and J. Rivi\`ere for the laboratory data. The seismic data used were obtained from the Plate Boundary Observatory \cite{PBO}. All the data is publicly available. This work was supported by DOE Office of Science (Geoscience Program, grant 89233218CNA000001) and Institutional Support (LDRD) at Los Alamos. CH work was also done thanks to a joint research laboratory effort in the framework of the CEA-ENS “Yves Rocard” LRC (France), and thanks to the European Research Council (ERC) under the European Union's Horizon 2020 research and innovation program (Geo-4D project, grant agreement 758210).

\section*{Author contributions statement}
C.H. and B.R.L. conducted the machine learning analysis. P.A.J. supervised the project. All authors contributed to writing the manuscript. 

\bibstyle{Science}
\bibliographystyle{ScienceAdvances}

\end{document}